\documentclass[aps,prd,twocolumn,amsmath,amssymb]{revtex4}
\usepackage{natbib}
\usepackage{graphicx}
\usepackage{dcolumn}
\usepackage{bm}

\begin{document} 

\title{Modified Weibull and Tsallis Distributions for multiplicities in $e^{+}e^{-}$ collisions at LEP2 energies}
\author{S. Sharma}
\author{M. Kaur}
\email{manjit@pu.ac.in} 
\author{S. Thakur}%
\affiliation{Department
of Physics, Panjab University, Chandigarh -160 014, India.}

\date{\today}

\begin{abstract}

Multiplicity distributions of charged particles produced in the $e^{+}e^{-}$ collisions at LEP2 energies ranging from 91 to 206 GeV  in full phase space, are compared with predictions from Tsallis $q$-statistics and the recently proposed Weibull distribution functions.~The analysis uses data from two LEP experiments, L3 and OPAL.~It is shown that Tsallis $q$-statistics explains the data in a statistically acceptable manner in full phase space at all energies, while the Weibull distribution fails to explain the underlying properties of the data.~Modifications to the distributions proposed earlier, are applied to uncover manifold improvements in explaining the data characteristics.

\end{abstract}  

\maketitle

\section{Introduction}
The particle production in high energy particle-particle interactions can be understood in terms of several theoretical and phenomenological models.~Several of these models are derived from quantum chromodynamics, laws of fluid mechanics, statistical mechanics, thermodynamics, hydrodynamics etc.~These models have been intriguingly successful in explaining the data trends.~To understand the production mechanism, concepts from ensemble theory in statistical mechanics have been used to develop models which include statistical fluctuations as an important source of information.~Several  distributions based on the statistical analyses have played an important role in the understanding of multiplicity distributions.~A host of such distributions can be found in the references \cite{NB1,NB2,KNO,MNBD,KW,LN,Wei}.~Several of the distributions such as Tsallis \cite{TS1,TS2}, Gamma \cite{Gam1}, HNBD \cite{Gam2} etc. have been derived from the concepts of statistical mechanics and ensemble theory.~The novel approach in the Tsallis $q$-statistics incorporating non-extensive entropy to describe the particle production has been successfully applied to heavy-ion and p-p collisions at some energies \cite{Zhang}.~The non-extensive property of the entropy is quantified in terms of a parameter $q$ which turns out to be more than unity under this assumption.~Weibull distribution is another statistical distribution which has been studied to describe multiplicity distributions in $e^{+}e^{-}$ and also for $ep$ collisions by authors S. Dash et al \cite{Wei} and S.Hegyi \cite{Heg1,Heg2}.

In the present study, we focus on investigating the multiplicity distributions, in full phase space, to study the characteristic properties of charged particle production in $e^{+}e^{-}$ collisions at different energies at LEP2. In one of our recent publications \cite{MK}, we analysed the data $\sqrt{s}$=14 to 91 GeV in terms of Tsallis distribution and Weibull distribution.~We also proposed a modification to improve the comparison between the predicted and the experimental values.~The modification of distributions was done by the convolution of PDFs from 2-jet fraction and multi-jet fractions by assigning appropriate weights. The 2-jet fractions for different energies were calculated from the DURHAM algorithm.~It was observed that the multiplicity distribution shows a shoulder structure at higher energies(Give reference).~The presence of a shoulder structure in the multiplicity distribution was observed at LEP1 energy of $\sqrt{s}$=91 GeV.~The multiplicity distributions at LEP2 energies confirmed the presence of shoulder structure at higher energies as well.~In our earlier publication \cite{MK} we had analysed the data only at 91 GeV and observed that the modification of distributions for both Tsallis and Weibull distributions improves the fitting by several orders, both in restricted rapidity windows and in full phase space.~However our results were not absolutely conclusive in the absence of analyses of data from higher energies.~In this paper we extend the analysis by comparing the multiplicity distributions obtained from Tsallis $q$-statistics with the Weibull distribution in the full phase space with the experimental results for $\sqrt{s}$ = 91 to 206 GeV at LEP2 to understand the constraints for the models used.~The data at LEP2 was not fully exploited which motivated us to carry out this analysis.

In Section II, we give a very brief outline of probability distribution functions (PDF) of Tsallis, Weibull and their modified forms along with the references for full details.~These details were given in our earlier publications also \cite{MK,SS}. But for the sake of completeness, these are outlined in this paper also.~The modification of distributions done by the convolution of PDFs from 2-jet fraction and multi-jet fractions by using appropriate weights, is implemented to improve the fitting results.~The 2-jet fractions for various energies have been taken from the DURHAM algorithm \cite{KT,DURHAM}, one of the most widely used algorithm for LEP data analyses.

~Section III presents the analyses of experimental data and the results obtained by two approaches.
~Discussion and conclusion are presented in Section IV. 

\section{Charged Multiplicity Distributions, Tsallis and Weibull}
Charged particle multiplicity is defined as the average number of charged particles, $n$ produced in a collision $<n>=\sum\limits_{n=0}^{n_{max}} {nP_n} $.~We briefly outline the distributions used for studying the multiplicity distributions;

\subsection{Tsallis and modified Tsallis distributions}
Tsallis statistics deals with entropy in the usual Boltzman-Gibbs thermo-statistics modified by introducing 
$q$-parameter and defined as;

\begin{equation}
S = \frac{1-\sum_{a}P_{a}^q}{q-1}\, \label{one}
\end{equation}
where  $P_a$  is the probability associated with microstate $a$ and sum of the probabilities over all microstates is normalized to one;
   $\sum_{a}P_{a}=1$.

 Tsallis entropy is defined as;
 \begin{equation}
 S_q(A,B)= S_A + S_B+ (1-q)S_{A}S_{B}
 \end{equation}
where $q$ is entropic index with value, $q>1$ and $1-q$ measures the departure of entropy from its extensive behaviour.

In Tsallis $q$-statistics probability is calculated by using the partition function Z, as  
\begin{equation}
P_N = \frac{Z^{N}_q}{Z}
\end{equation}
where Z represents  the total partition function and $Z^{N}_q$ represents partition function at a particular multiplicity.

For N particles, partition function can be written as, 
\begin{equation}
Z(\beta,\mu,V) = \sum(\frac{1}{N!})n^{N}(V-Nv_{0})^{N}\Theta(V-Nv_{0})
\end{equation}
$n$ represents the gas density, V is the volume of the system and $v_{0}$ is the excluded volume associated to a particle.~The Heavyside $\Theta$-function limits the number of particles inside the volume $V$ to $N < V/v_{0}$.~ $\bar{N}$, the average number of particles, is given by 
 \begin{equation}
 \bar{N}=Vn[1 + (q-1)\lambda(V n\lambda-1)- 2v_0n] 
 \end{equation} 
where $\lambda$ is related to the temperature through the parameter $\beta$ as;
\begin{equation}
\lambda(\beta,\mu)= -\frac{\beta}{n}\frac{\partial n}{\partial\beta}
\end{equation} 
 $K$-parameter is related to $q$ and excluded volume, by
 \begin{equation}
\frac{1}{K}=(q-1)\lambda^{2} - 2\frac{v_0}{V} 
\end{equation} 
Details of the Tsallis distribution and how to find the probability distribution can be obtained from \cite{TS2}.~In one of our earlier papers, we have analysed the $e^{+}e^{-}$ interactions at various energies for full phase space data and described the procedure in detail in reference \cite{SS}.~We also proposed to modify and build the multiplicity distribution by adding weighted superposition of multiplicity in 2-jet events and multiplicity distribution in multi-jet events.~We then calculated the probability function from the weighted superposition of Tsallis distributions of these two components, as given below; 

\begin{multline}
P_{N}(\alpha:\bar{n_1},V_1,v_{01},q_1:\bar{n_2},V_2,v_{02},q_2)= \\
\alpha P_{N}(\bar{n_1},V_1,v_{01},q_1) +\\
(1-\alpha)P_{N}(\bar{n_2},V_2,v_{02},q_2)
\end{multline} 

where $\alpha$ is a weight factor which gives 2-jet fraction from the total events and is determined from a jet finding algorithm.
 
\subsection{ Weibull and modified Weibull distributions} 

 Weibull distribution is a continuous probability distribution which can take many shapes. It can also be fitted to non-symmetrical data. 
 
The probability density function of a Weibull random variable is;
\begin{multline}
 P_N(N,\lambda,k) = \Bigg\lbrace\frac{k}{\lambda} (\frac{N}{\lambda})^{(k-1)} exp^{-(\frac{N}{\lambda})^{ k}} \hspace{0.8cm} N \geq 0 \\                                   
0  \hspace{3.6cm}  N < 0 
\end{multline} 
 
The standard Weibull has characteristic value $\lambda >0$, also known as scale factor, and shape parameter $k>0$ for its two parameters.The two parameters for the distribution are related to the mean of function, as 
\begin{equation}
\bar{N} = \lambda \Gamma(1+1/k)
\end{equation}

Modified Weibull distribution has been obtained by the weighted superposition of two Weibull distributions to produce the multiplicity distribution.~We convolute the weighted distributions due to 2-jet component and multi-jet component of the events, as below; 

\begin{multline}
P_{N}(\alpha:N_1,\lambda_1,k_1;N_2,\lambda_2,k_2)=\\
\alpha P_{N}(N_1,\lambda_1,k_1) + \\
(1-\alpha)P_{N}(N_2,\lambda_2,k_2) 
\end{multline}

where $\alpha$ is the weight factor for 2-jet fraction out of the total events and the remaining $1-\alpha$ is the multi-jet fraction.~$\alpha$ is calculated from the DURHAM jet algorithm, as discussed in the next section.

\section{Analysis on Experimental data \& Results}

Experimental data on $e^{+}e^{-}$ collisions at different collision energies at LEP2 and from two experiments are analysed.~Details of the data used from the experiments, L3 \cite{L3} and OPAL \cite{OPAL91, OPAL133, OPAL161, OPAL172} at different energies between $\sqrt{s}$ = 91 to 206 GeV in the full phase space are given in Table I. ~The experimental distributions are fitted with the predictions from Tsallis $q$-statistics and the Weibull distributions as described in the following two sections.
 
\subsection{Tsallis versus Weibull}
The probability distributions using Tsallis distribution function and Weibull function are calculated using equations (3-7) \& (9-10) and fitted to the experimental data.~Figure 1 shows the Tsallis fits to the data and figure 2 shows the Weibull distributions fitted to the data in  different center of mass energies  for L3 and OPAL data.

We find that overall, Weibull fails to reproduce the distributions, particularly in the high multiplicity intervals, while Tsallis distribution shows good fit in full phase space.~Detailed comparison between the two functions is shown in Tables~II \& VI where $\chi^{2}/ndf$
and $p$ values at all energies are given.~It is observed that the  $\chi^{2}/ndf$ values are considerably lower for the Tsallis fittings in comparison to the Weibull fittings.~This is true for all energies.~A careful examination of the p-values shows that for the data from L3 experiment at all energies from 130.1 to 206.1 GeV, Weibull fits  are statistically excluded with $CL < 0.1 \%$.~While for Tsallis fits, the data only at 200.1 and 206.1 are statistically excluded with $CL < 0.1 \%$ and is good for all other energies with $CL > 0.1 \%$. 
~For the data from OPAL experiment, Weibull fit is statistically excluded systematically for all energies between 91 to 189 GeV with $CL < 0.1 \%$ except being good for energy 172 GeV.~Again Tsallis fit is excluded only for one energy at 91 GeV and remains a good fit for all energies from 131 to 189 GeV with $CL > 0.1 \%$.

A comparison of the $\chi^{2}/ndf$ values in Tables~II \& VI reveals that $\chi^{2}/ndf$ values for the Tsallis distributions are lower by several orders, confirming that Tsallis distribution fits the data far better than Weibull.

\begin{figure}

\includegraphics[width=3.3 in, height =2.3 in]{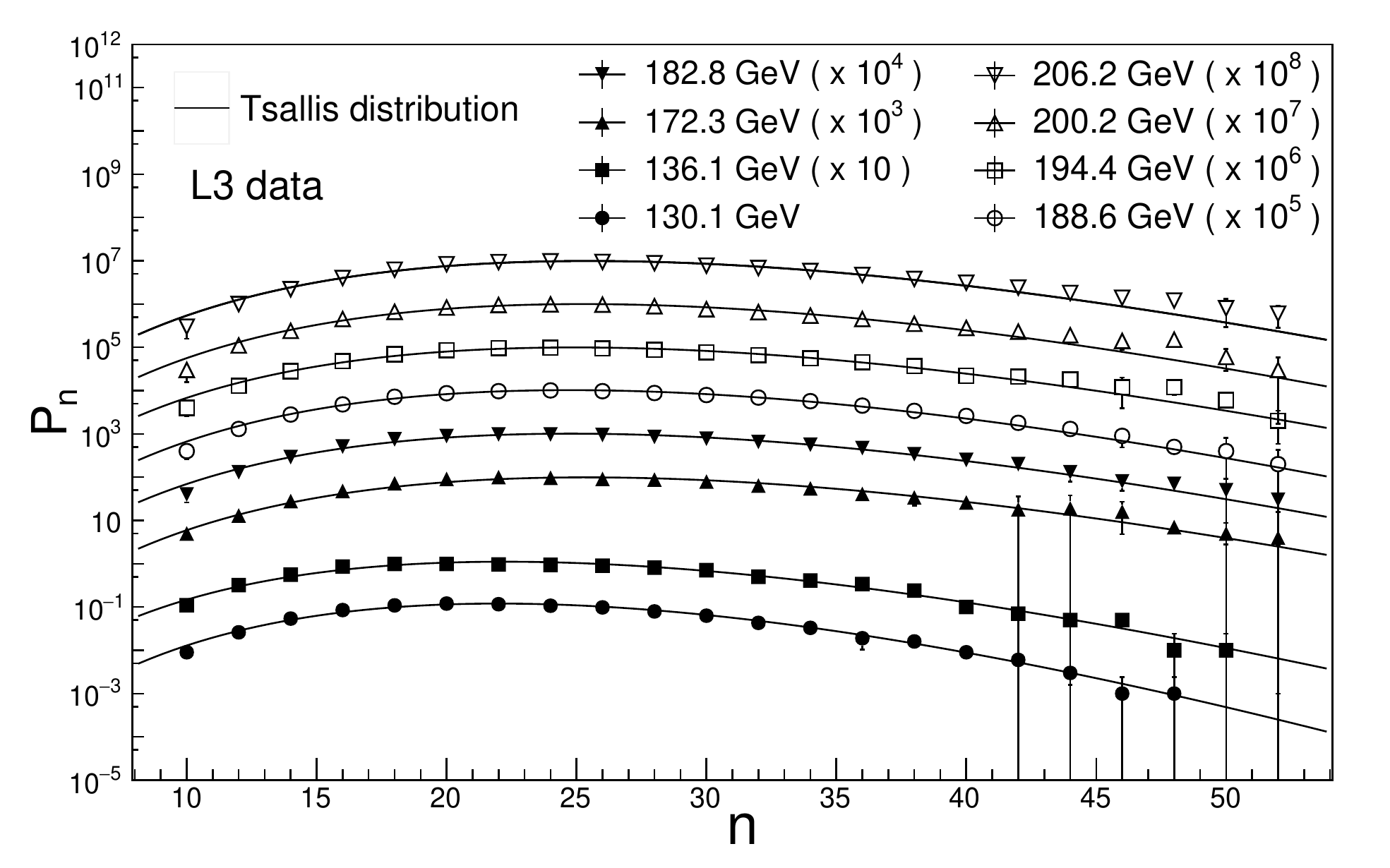}
\includegraphics[width=3.3 in, height =2.3 in]{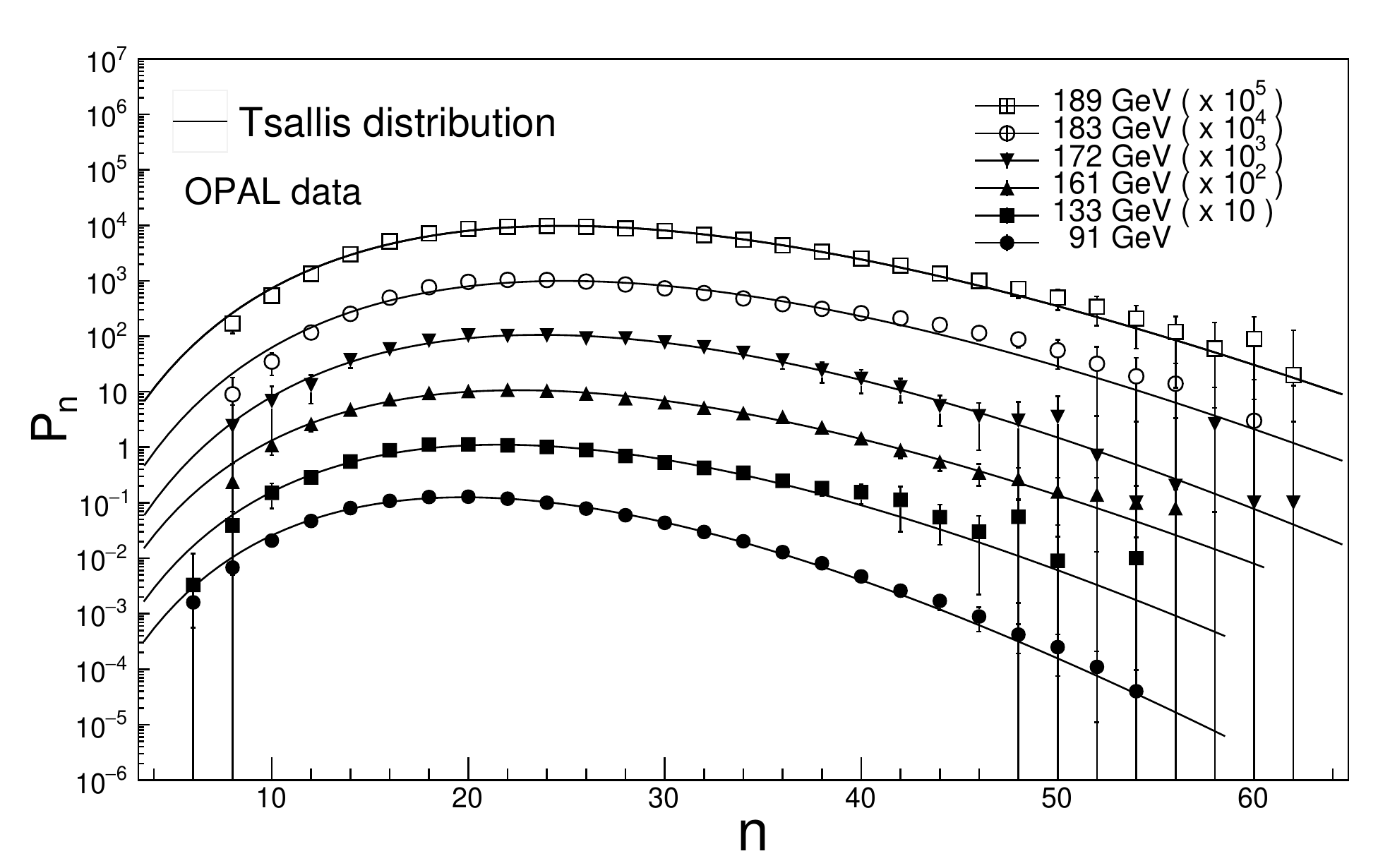}
\caption{Charged multiplicity distribution from top to bottom, $\sqrt{s}$ = 206 to 130 GeV from the L3 and OPAL experiments at LEP2.~Solid lines represent the Tsallis distribution and points represent the data in the two plots.}
\end{figure}

\begin{figure}
\includegraphics[width=3.3 in, height =2.3 in]{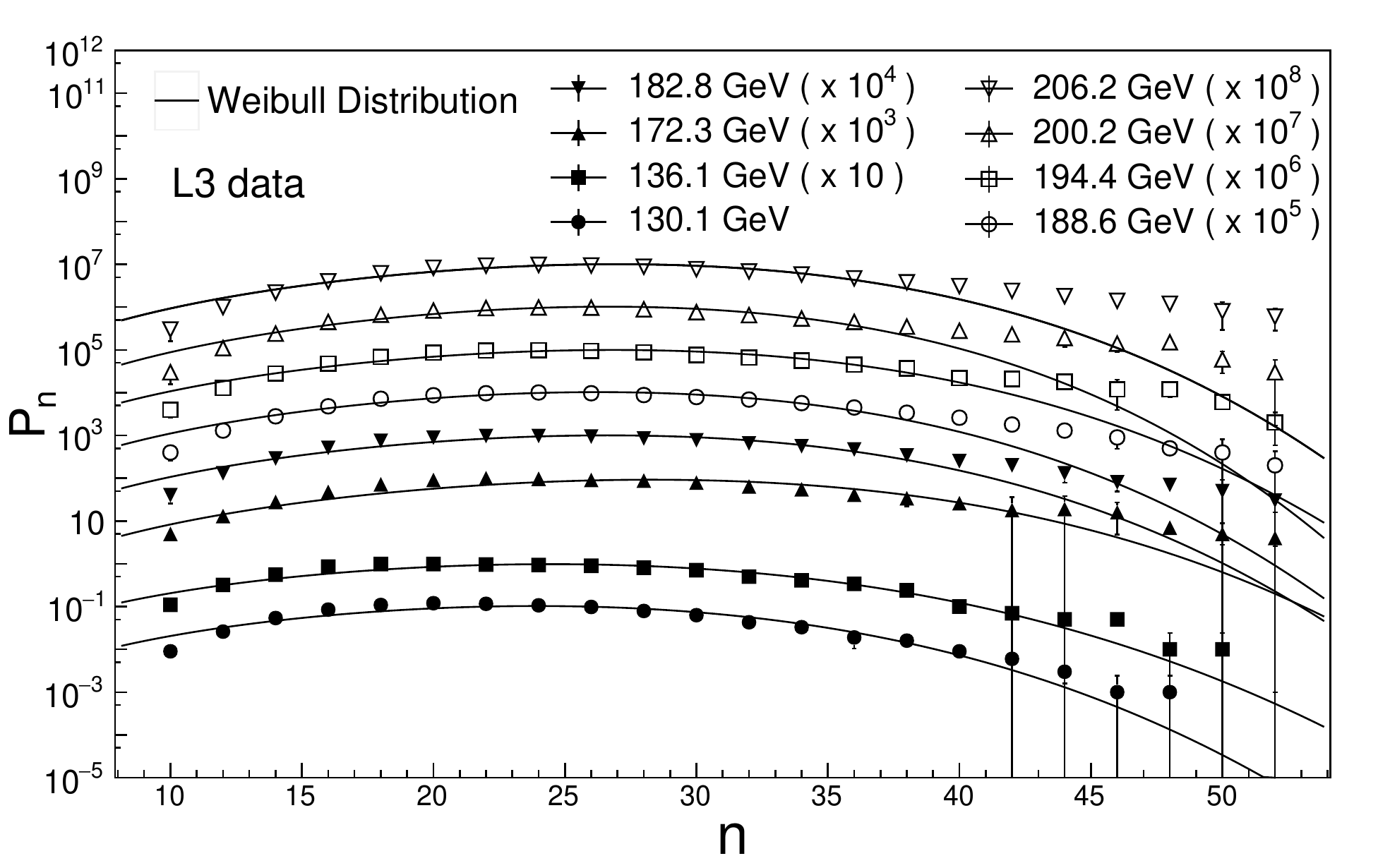}
\includegraphics[width=3.3 in, height =2.3 in]{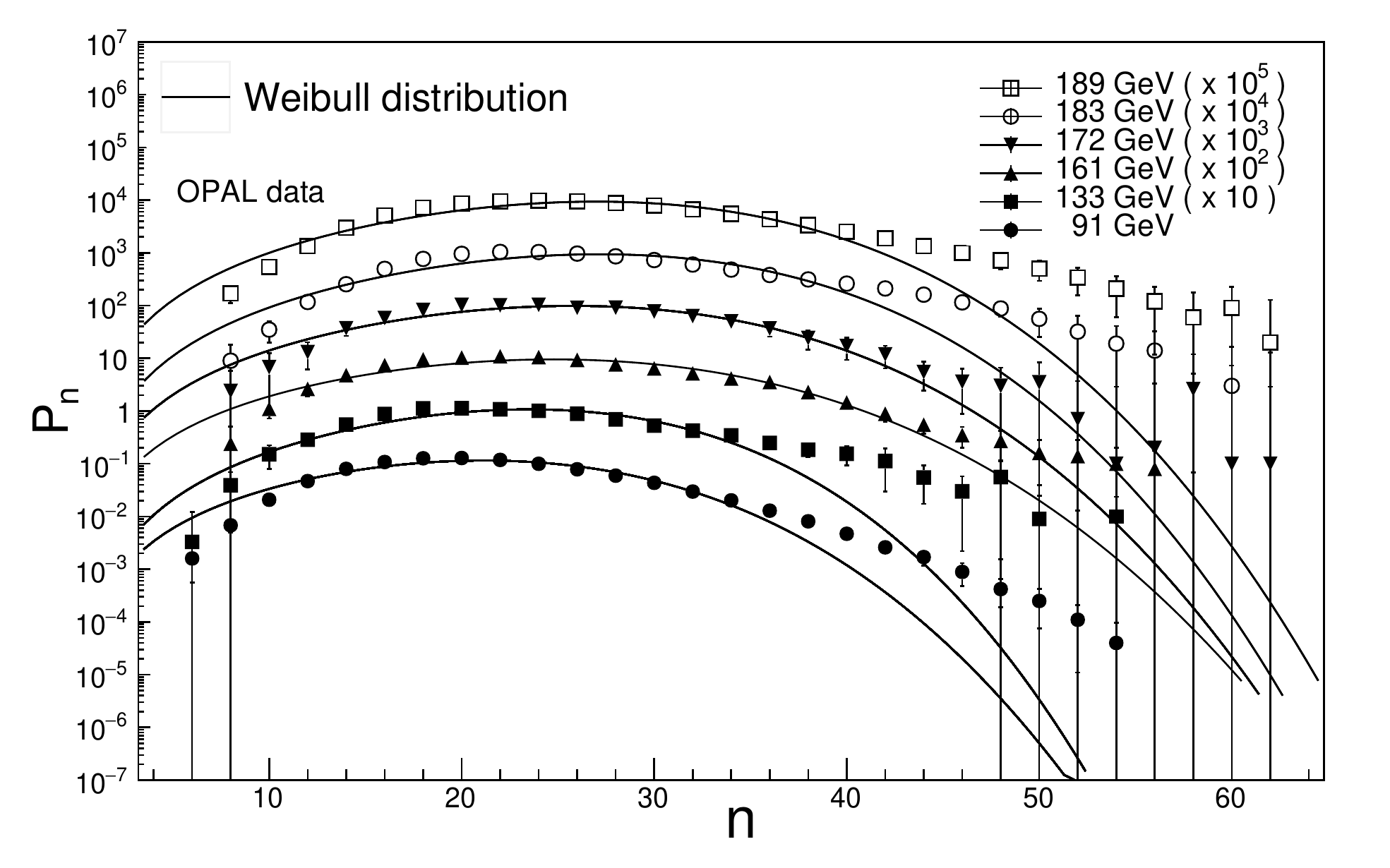}
\caption{Charged multiplicity distribution from top to bottom, $\sqrt{s}$ = 206 to 130 GeV from the L3 and OPAL experiments at LEP2.~Solid lines represent the Weibull distribution and points represent the data in the two plots.}
\end{figure}

\begin{figure}
\includegraphics[width=3.3 in, height =2.3 in]{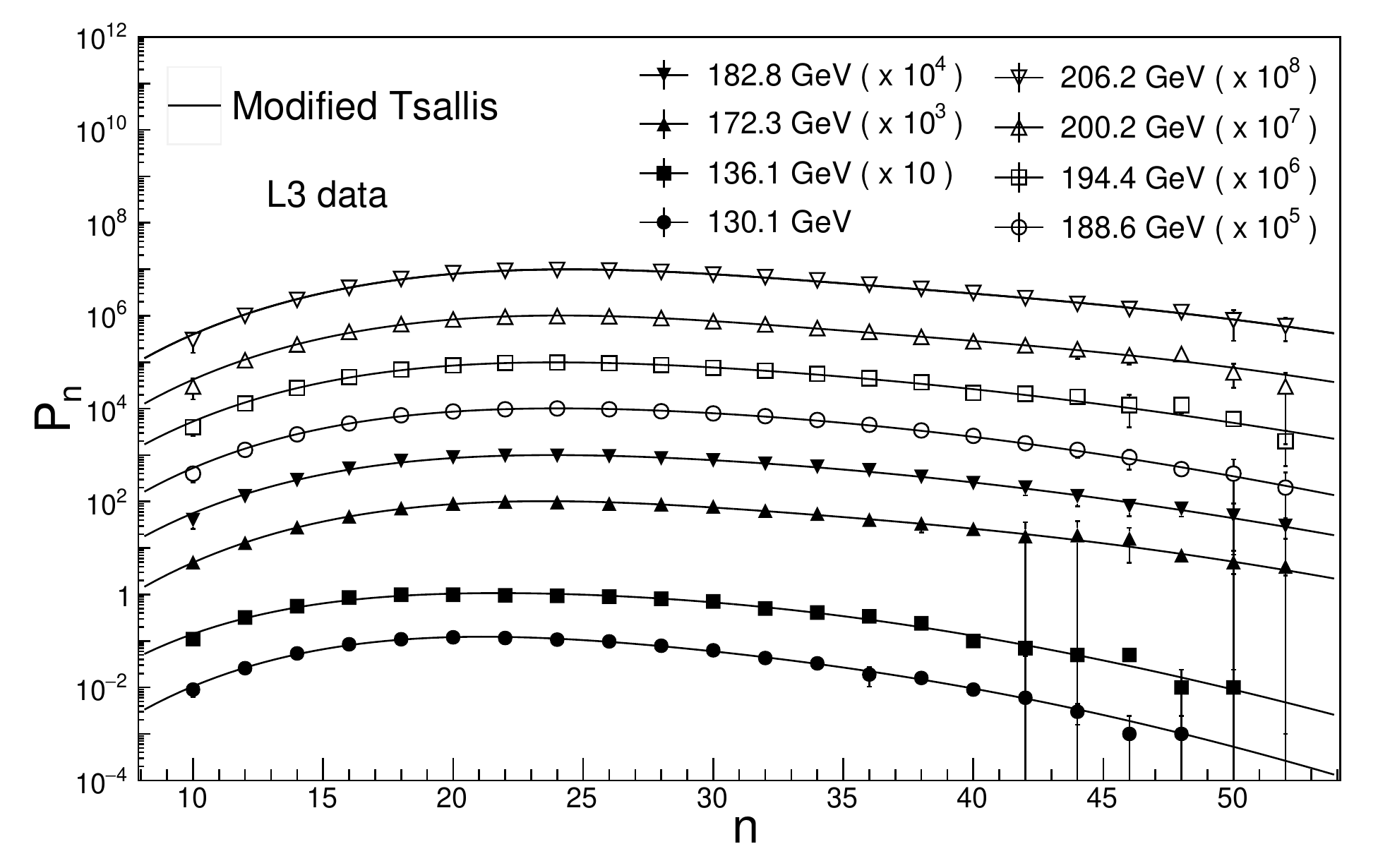}
\includegraphics[width=3.3 in, height =2.3 in]{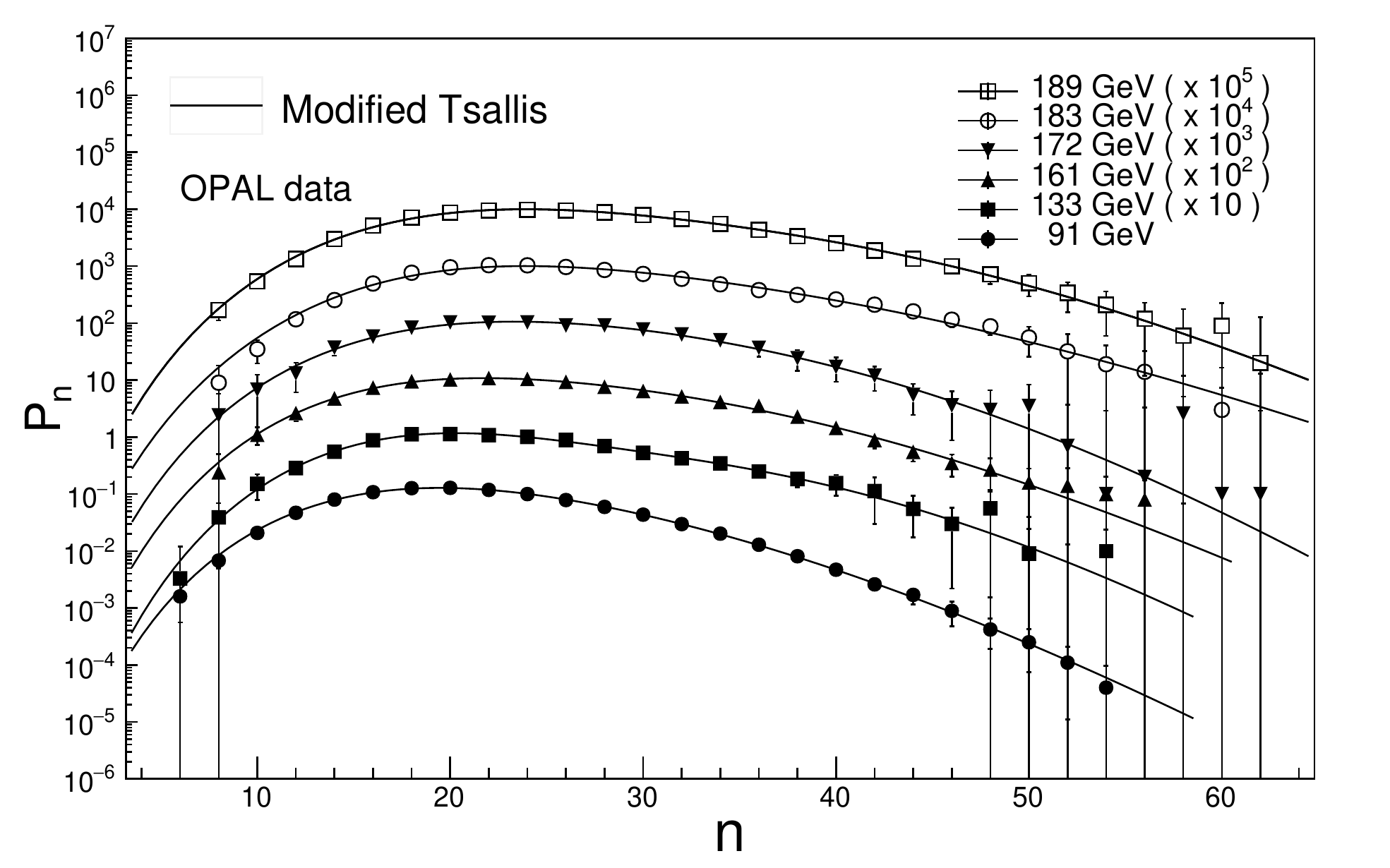}
\caption{Charged multiplicity distribution from top to bottom, $\sqrt{s}$ = 206 to 130 GeV from the L3 and OPAL experiments at LEP2.~Solid lines represent the modified Tsallis distributions and points represent the data in the two plots.}
\end{figure}

\begin{figure}
\includegraphics[width=3.3 in, height =2.3 in]{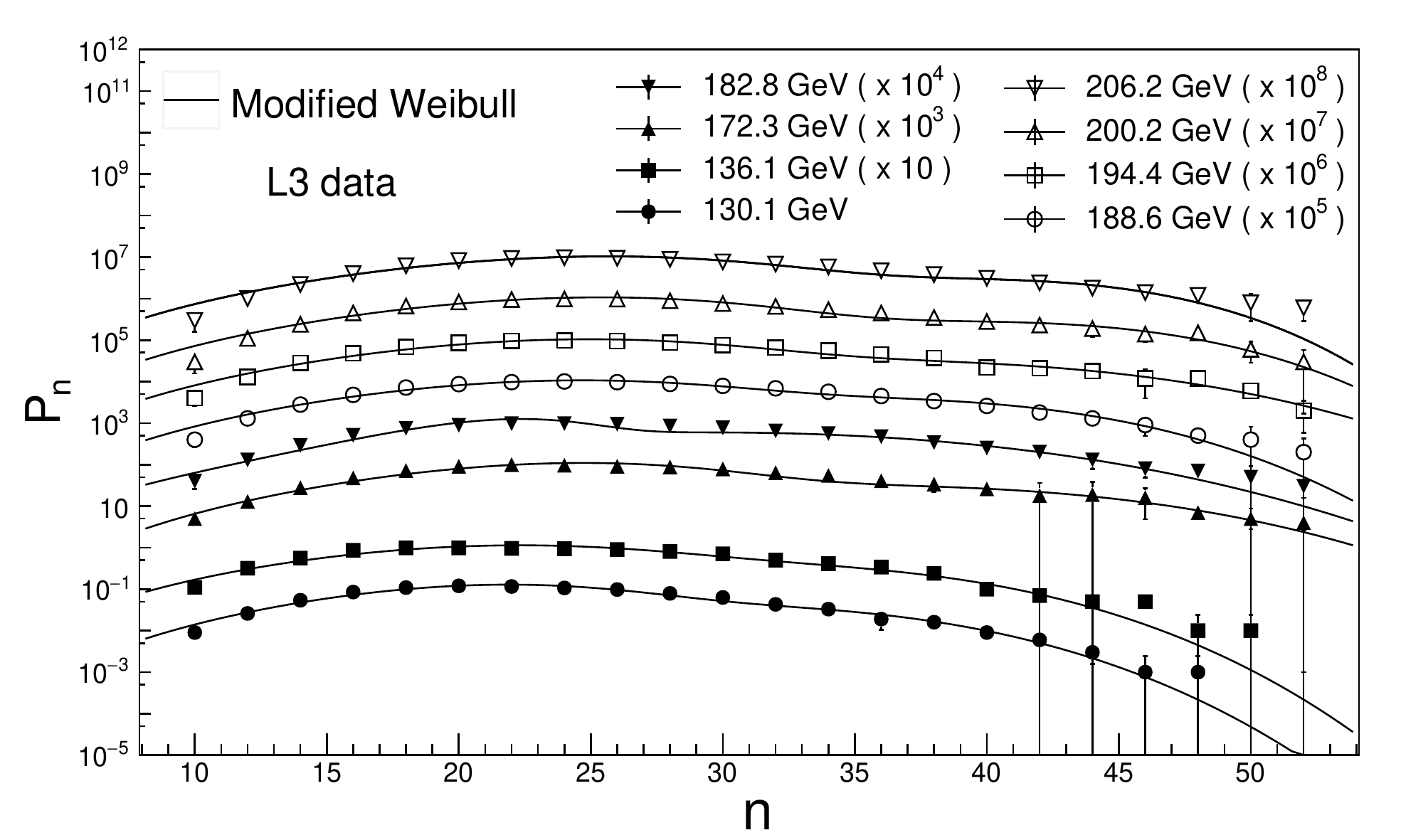}
\includegraphics[width=3.3 in, height =2.3 in]{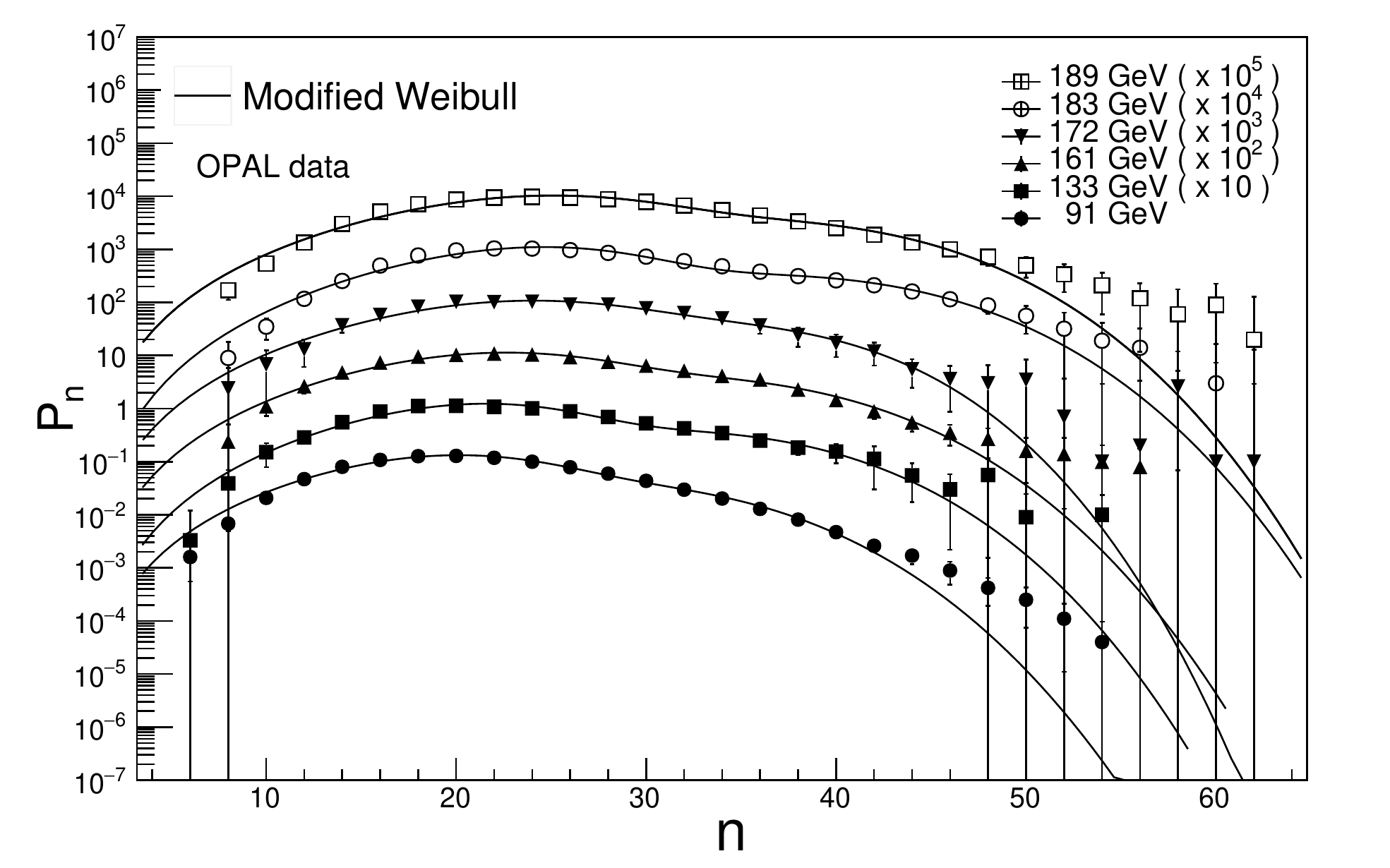}
\caption{Charged multiplicity distribution from top to bottom, $\sqrt{s}$ = 206 to 130 GeV from the L3 and OPAL experiments at LEP2.~Solid lines represent the modified Weibull distribution and points represent the data in the two plots.}
\end{figure}

 \begin{figure}
\includegraphics[width=3.3 in, height =2.3 in]{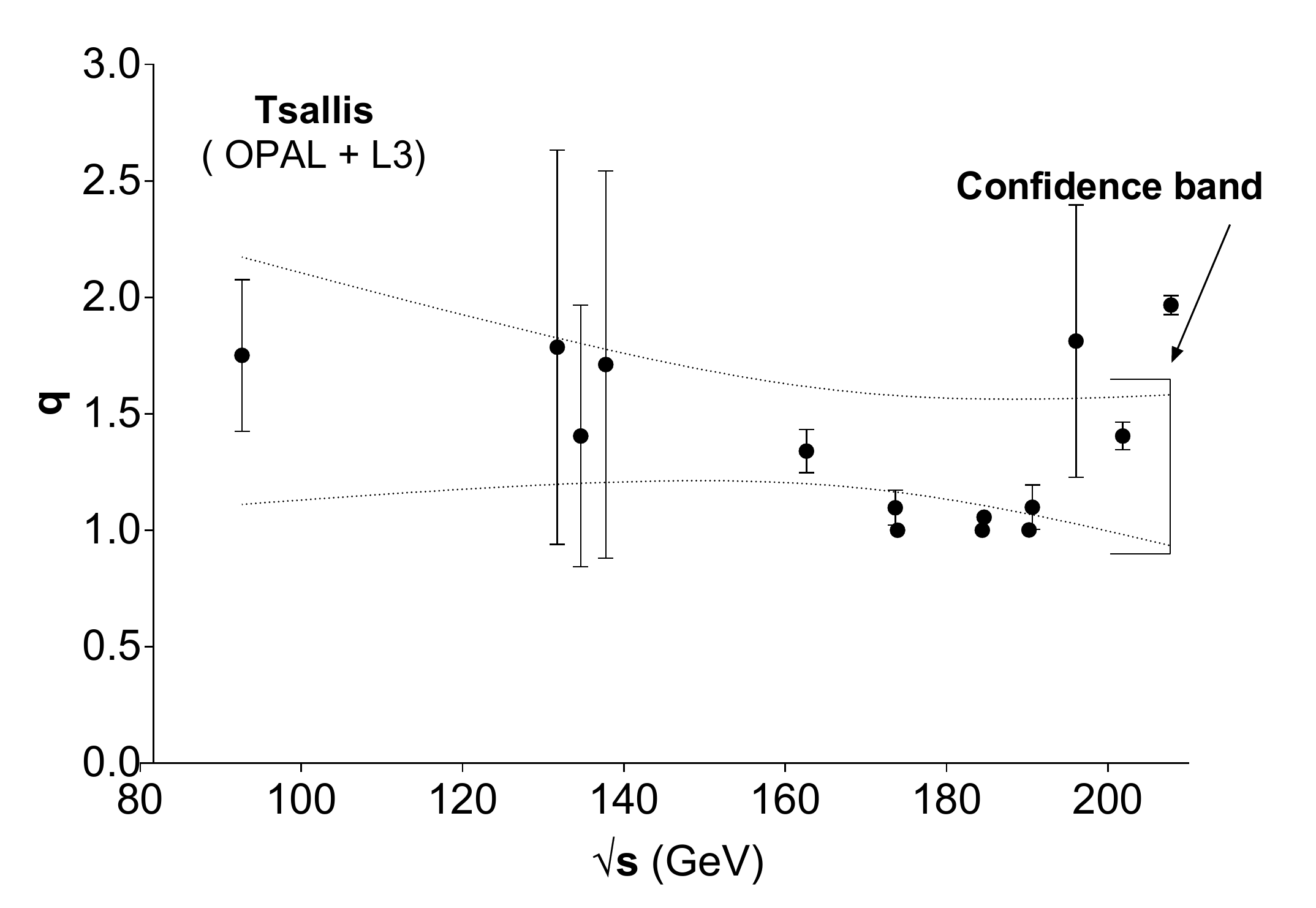}
\includegraphics[width=3.3 in, height =2.3 in]{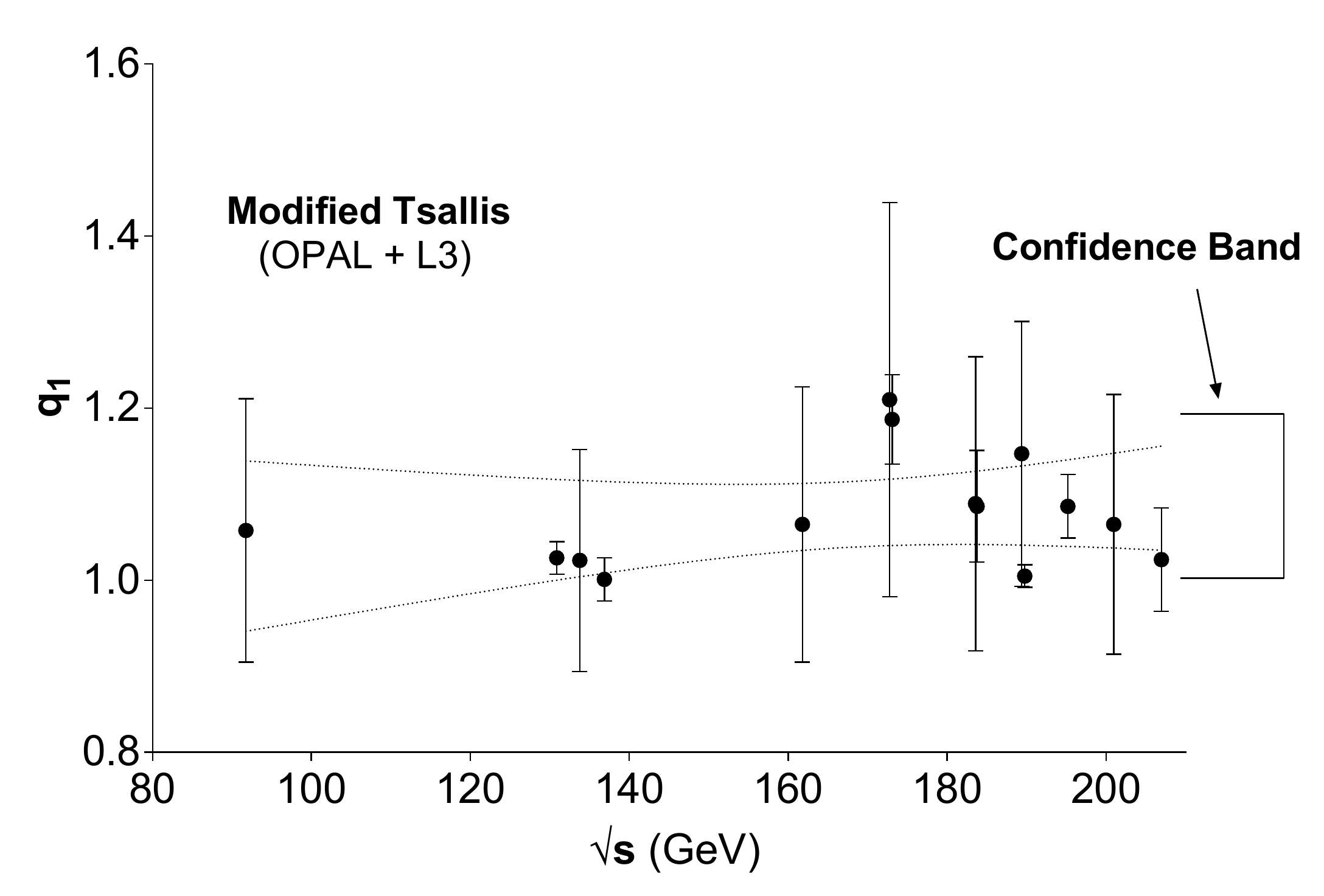}
\includegraphics[width=3.3 in, height =2.3 in]{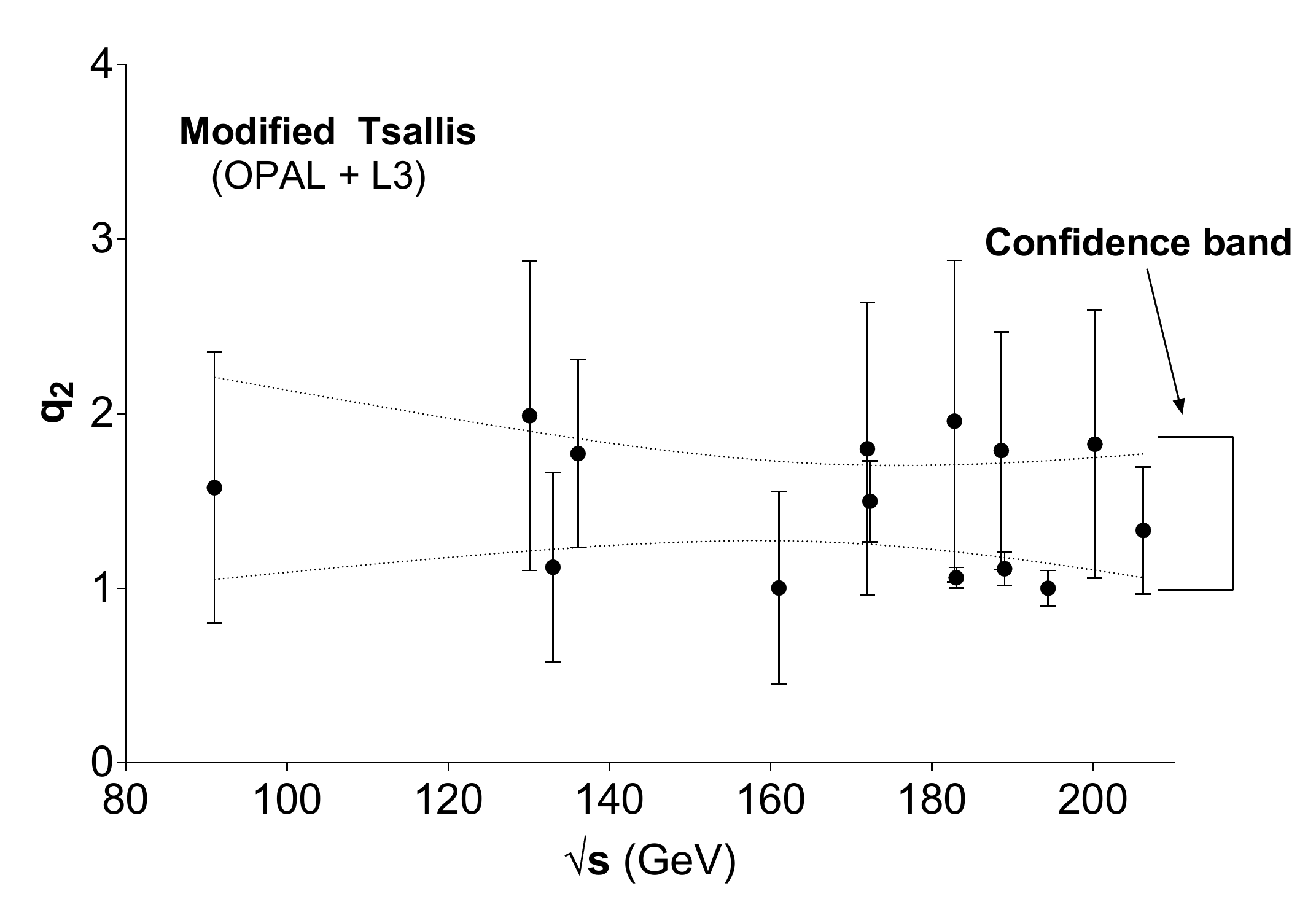}
\caption{$q$, $q1$ \& $q2$ values for $\sqrt{s}$=91 to 206 GeV from the L3 and OPAL experiments at LEP2.~Points represent the data and the bands shown are confidence bands.}
\end{figure}
 
~It is also observed that for Weibull distribution, $\lambda$ values increase with energy, as expected.~Similarly for Tsallis distribution, the $q$ value which measures the entropic index of the Tsallis statistics, is more than 1 in every case.~This confirms the non-extensivity of the Tsallis statistics. 


\subsection{Modified Tsallis versus Modified Weibull} 

It was observed \cite{Gov} that the multiplicity distributions have a shoulder-like structure at high energies.~The Tsallis and the Weibull distributions both give very high $\chi^{2}/ndf$ values and do not describe the data well at high energy.~In our previous publication \cite{SS} we suggested to adopt the Giovannini's approach \cite{Gov} whereby the multiplicity distribution is obtained by using the weighted superposition of two distributions; one accounting for the 2-jet events and another for multi-jet events.~For the present work, we use this approach on both Tsallis and Weibull distributions.~We call these as modified Tsallis and modified Weibull distributions.~The probability functions for the two cases are  given in equations (8 \& 11).~Using the modified distributions, data at all energies mentioned in TableI have been analysed.The 2-jet fraction used in the analysis are derived from the DURHAM algorithm, as explained in references \cite{DURHAM, SD}.\\ 

Using this approach for $\sqrt{s}$= 130 to 206 GeV, data from L3 and OPAL Collaborations are analysed for full phase space.~We use these data as the shoulder structure is prominent at this energy.~Fit parameters, $\chi^{2}/ndf$ and $p$ values for both modified Weibull and modified Tsallis distributions are given in Table~VI.~Figure 3 \& 4 show the comparison of distributions for L3 and OPAL data.~It may be observed from these two figures and the Tables III \& IV that by using this approach, the fits to the data improve enormously and the $\chi^{2}/ndf$ values decrease substantially.~Both the Modified Weibull distribution and the modified Tsallis distribution describe well, the data at various energies.~However, Weibull fails for energies 161,188.6, 189 \& 206.1 GeV where $CL < 0.1 \%$.

~The fit procedure uses ROOT 5.36 from CERN to minimise the $\chi^{2}$ using the library MINUIT2.~The fitting procedure for Tsallis distribution involves 4 free parameters and a normalisation constant.~Several options of minimization (MIGRAD, MINOS, FUMILI) had to be tried to get the covariance matrix positive definite.~Also $n$,$V$ and $v_{0}$ are correlated, we had to choose $nV$ and $nV_{0}$ as free parameters for meaningful fits independent of the starting values of fit parameters.~In the Weibull distribution, the fit procedure is more straight forward without involving such constraints.~However, for both modified Tsallis and modified Weibull distributions, the fit parameters are doubled while introducing the modification.~Minimization then leads to larger errors on the fit parameters, especially in $K2$ $\&$ $q2$ for Tsallis and $\lambda2$ $\&$ $K2$ for Weibull.~This may lead to very large $p$ values, particularly close to unity.~The additional uncertainty comes in because of very low statistics at some energies.
~Figure 5 shows plots of the $q$, $q1$ and $q2$ values estimated from Tsallis and modified Tsallis fits.~The band shown are the confidence bands.~The mean values are $q=1.3879 \pm 0.0950$, $q1=1.0766 \pm 0.0174$ and $q2=1.48864 \pm 0.0998$.~The figure shows that the q-values in all cases, Tsallis and modified Tsallis, are more than unity, emphasising the non-extensive nature of entropy. 

\section{CONCLUSION}
Detailed analysis of the data on $e^{+}e^{-}$ collisions at LEP2 energies, $\sqrt{s}$=130 to 206 GeV has been done by fitting the recently proposed Weibull distribution in comparison to the Tsallis distribution.~For the sake of consistency cross check with earlier results, we have also considered 91 GeV from OPAL.~It is observed that the Weibull fits for data from L3 experiment at all energies from 130.1 to 206.1 GeV are statistically excluded with $CL < 0.1 \%$.~While for Tsallis fits the data only at 200.1 and 206.1 are statistically excluded with $CL < 0.1 \%$ and is good for all other energies with $CL > 0.1 \%$. 

~For the data from OPAL experiment, Weibull fit is statistically excluded for all energies between 91 to 189 GeV with $CL < 0.1 \%$, with an exception, is good only for energy at 172 GeV.~In contrast, Tsallis fit is excluded only for one energy at 91 GeV and remains a good fit for all energies from 131 to  189 GeV with $CL > 0.1 \%$.

A comparison of the $\chi^{2}/ndf$ values in Tables~II shows that $\chi^{2}/ndf$ values for the Tsallis distributions are lower by several orders than the Weibull distribution, confirming that Tsallis distribution fits the data far better than Weibull.

The shape parameter $k$ in the Weibull distribution affects the shape of the distribution.~Within the LEP2 energy range, the value of $k$ decreases slightly within limits of errors  with increasing energy, as can be seen in Tables II \& III.~This behaviour is related to the soft gluon emission and subsequent hadronization.~The scale parameter $\lambda$ of the Weibull distribution measures the width of the distribution.~Larger the scale parameter, more spread out the distribution is.~At higher collision energies, the mean multiplicity increases and so does the number of high multiplicity events.~As a result, $\lambda$ is expected to increase to account for the broader shape.~Similar results can be observed from the modified Weibull fit distribution parameters in table~III $\&$ VI where $\lambda1$ and $\lambda2$ values increase systematically from lower to higher energy.
In the Tsallis distribution, $K$ parameter measures the deviation from Poisson distribution and is related to the variance.~The definition of $K$ is motivated by the $k$ parameter of a negative binomial distribution, as given by equation (6).~Tsallis statistics for $q > 1$ with excluded volume $v_0$ produces the multiplicity distributions that are wider than the Boltzmann-Gibbs ones.~In some analysis the excluded volume is fixed between 0.3-0.4 $fm^{3}$. The corresponding value of volume $V$ then varies from few $fm^{3}$ to few tens of $fm^{3}$ \cite{TS2}. 

It is known that the multiplicity distributions at higher energies show a shoulder structure.~In order to improve upon both Weibull and Tsallis fits to the data, we proposed to build the multiplicity distribution by a convolution of 2-jet component and the multi-jet component.~For the various energies, the 2-jet fraction values calculated from various jet algorithms are available.~We show that by appropriately weighting the multiplicity distribution with the 2-jet fraction obtained from DURHAM algorithm for $\sqrt{s}$ = 91 to 206 GeV, both Tsallis and Weibull distributions describe the data well, giving the statistically significant results.~The modified Tsallis distribution reproduces the data well with $CL > 0.1 \%$ for all energies, from both L3 and OPAL experiments.~The modified Weibull distribution, also improves the fits by several orders, but fails to describe the data at most of the energy points, as observed from $p$ values in Table VI.~It gives good results only at 130.1 and 172.3 GeV.

~In the fitting of the PDFs, modifide Weibull fits have 4 free parameters while modified Tsallis distribution has 8 free parameters.~Due to limited number of data points, fit parameters suffer from large errors, especially in Tsallis distribution.~While the Tsallis has the disadvantage of 
larger number of fit parameters, Weibull offers a simplistic approach with fewer parameters.~Nevertheless the present detailed analysis establishes that the performance of Tsallis is superior to the Weibull.~The $q$ value known as entropic index in Tsallis distribution, accounts for the nonextensive thermostatistical effects in hadron production and is expected to be more than 1.~The mean values of $q$ measured for total sample of events, events with two jets and events with multijets are $q = 1.3879 \pm 0.0950$ , $q1=1.07657 \pm 0.1737$ and $q2 = 1.48864 \pm 0.0998$ respectively.~This not only confirms the non-extensive behaviour of Tsallis entropy but also asserts that the behaviour is more pronounced in the events with higher multiplicity.~Thus, the analysis presented, using the data at higher energies from different experiments asserts our conclusion presented earlier, on the comparison between Tsallis and Weibull fits.

\begin{table*}[t]

\begin{tabular}{|c|c|c|c|c|}
\hline
Energy  & Collaboration & No. of events  & $\alpha$ values & Reference  \\\hline

(GeV) & &  &  &  \\\hline
 
91 & OPAL & 82941    & 0.657 & \cite{OPAL91,DURHAM}\\\hline
133 & ''  & 13665    & 0.662 & \cite{OPAL133,DURHAM} \\\hline
161 & ''  & 1336     & 0.635 & \cite{OPAL161,DURHAM} \\\hline
172 & ''  & 228      & 0.666 & \cite{OPAL172,DURHAM,MA}\\\hline
183 & ''  & 1098     & 0.675 & \cite{OPAL1721} \\\hline
189 & ''  & 3277     & 0.662 & '' \\\hline

 & &  &   &  \\
130.1 & L3 & 556  & 0.654 & \cite{L3} \\\hline

136.1 & ''  & 414   & 0.649 & ''\\\hline

172.3 & ''  & 325    &0.657 & ''\\\hline
182.8 & ''  & 1500 & 0.668 & ''\\\hline

188.6 & ''  & 4479   & 0.670 & ''\\\hline

194.4 & ''  & 2403  & 0.679 &''\\\hline
200.2 & ''  & 2456    & 0.661 & ''\\\hline
206.2 & '' & 4146   & 0.666 & '' \\\hline

\end{tabular}
\caption{Experimental data for the charged multiplicity} 
\end{table*}

\begin{table*}[t]
\begin{tabular}{|c|c|c|c|c|c|c|c|c|}
\hline
     &         &  			  & &          &                &   &  & \\
  & Weibull &$\rightarrow$ & &  Tsallis  &$\rightarrow$   &    & & \\\hline       
  Energy & k  & $\lambda$ & $\chi^{2}/ndf$  & $nV$ & $nv_{0}$ & K & q & $\chi^{2}/ndf$ \\
  (GeV)	 &	  &			 &					 &	  &   &    &   &\\\hline    
 
91  &	3.548  $\pm$  0.033 &	23.197  $\pm$  0.073	& 434.91/22 & 13.104  $\pm$  0.312	& 0.115  $\pm$  0.325 &	21.397  $\pm$  0.670	& 1.751  $\pm$  0.325 & 33.13/20\\\hline

131  & 4.029  $\pm$  0.122	& 25.239  $\pm$  0.374 & 66.88/22 & 13.303  $\pm$  0.220 &	0.124  $\pm$  0.169 &	19.891  $\pm$  2.251	& 1.405  $\pm$  0.562	& 12.67/20 \\\hline

161  & 3.542  $\pm$  0.098 &	27.174  $\pm$  0.300 & 48.11/22 & 13.244  $\pm$  0.178 &	0.134  $\pm$  0.005	& 17.983  $\pm$  1.587 & 1.340  $\pm$  0.093 & 5.81/20 \\\hline

172  & 3.813  $\pm$  0.162 & 27.942  $\pm$  0.479	& 17.87/22 & 13.511  $\pm$  0.199 &	0.164  $\pm$  0.011	& 21.274  $\pm$  3.003 &	1.097  $\pm$  0.075	& 3.94/20 \\\hline

183  & 4.069  $\pm$  0.088 & 29.117  $\pm$  0.277 & 159.61/25 & 13.523  $\pm$  0.065 &	0.179  $\pm$  0.007 &	19.749  $\pm$  1.339 & 1.056  $\pm$  0.023 & 35.13/23 \\\hline

189  & 3.994  $\pm$  0.058 & 29.018  $\pm$  0.197 & 183.63/25 & 13.415  $\pm$  0.305 &	0.167  $\pm$  0.010	& 17.795  $\pm$  0.858 & 1.099  $\pm$  0.096 & 15.88/23 \\\hline

	 &	  &			 &					 &	  &   &    &   &\\
130.1  & 3.655  $\pm$  0.095 &	26.110  $\pm$  0.215 &	52.53/19 &	12.241  $\pm$  0.979 &	0.338  $\pm$  0.103 &	23.513  $\pm$  2.031 &	1.786  $\pm$  0.846 &	6.42/17 \\\hline

136.1  &	3.513  $\pm$  0.099 &	26.913  $\pm$  0.235 &	41.64/19 &	12.004  $\pm$  0.947 &	0.306  $\pm$  0.109 &	18.309  $\pm$  1.490 &	1.712  $\pm$  0.831 &	19.26/17 \\\hline

172.3	 & 4.033  $\pm$  0.121 &	30.358  $\pm$  0.296 &	64.28/19 &	13.393  $\pm$  0.107	& 0.168 $\pm$   0.010	& 18.813  $\pm$  1.324 & 1.0001  $\pm$  0.0005 &	8.78/17 \\\hline

182.8	 & 4.041  $\pm$  0.101 &	28.590  $\pm$  0.358 &	98.71 / 19 &	13.544  $\pm$  0.070 &	0.132  $\pm$  0.026 &	18.983  $\pm$  1.332 &	1.0001  $\pm$  0.0001 &	16.98/17 \\\hline

188.6	 & 4.066  $\pm$  0.078 &	28.144  $\pm$  0.251 &	157.54/19 &	13.330  $\pm$  0.087 &	0.164  $\pm$  0.006 &	19.883  $\pm$  1.044 &	1.001  $\pm$  0.0006 &	17.28/17 \\\hline

194.4	 & 3.998  $\pm$  0.084 &	28.841  $\pm$  0.266 &	108.70/19 &	13.931  $\pm$  9.123 &	0.221  $\pm$  0.197 &	18.533  $\pm$  1.223 &	1.812  $\pm$  0.585 &	19.19/17 \\\hline

200.2	 & 4.271  $\pm$  0.099	& 28.350  $\pm$  0.322 & 127.53/19 &	14.511  $\pm$  0.094 &	0.158  $\pm$  0.004 &	20.092  $\pm$  1.497 &	1.405  $\pm$  0.060 &	27.14/17 \\\hline

206.2	 & 4.151  $\pm$  0.079 &	28.831  $\pm$  0.262 &	168.91/19 &	15.412  $\pm$  0.212	& 0.122  $\pm$  0.005 &	19.631  $\pm$  1.195 &	1.967  $\pm$  0.041 &	32.41/17 \\\hline

\end{tabular}
\caption{Parameters of Weibull and Tsallis functions for OPAL $\&$ L3 Experiment }  
\end{table*}

\begin{table*}[t]
\small
\begin{tabular}{|c|c|c|c|c|c|}
\hline
Energy   & $k_{1}$ & $\lambda_{1}$ & $k_{2}$ & $\lambda_{2}$  & $\chi^{2}/ndf$   \\
 & 	  			 		& &	&	 &	       \\\hline    
91  & 4.556  $\pm$  0.228 & 23.030  $\pm$  0.471 & 4.684  $\pm$  0.482 & 32.810  $\pm$  0.544	& 12.20/20	 \\\hline

133 &	4.810  $\pm$  0.192 &	22.050  $\pm$  0.418 &	4.811  $\pm$  0.557 &	32.480  $\pm$  0.775	& 14.71/20  \\\hline

161  &	4.412  $\pm$  0.063 &	20.620  $\pm$  0.119 &	4.221  $\pm$  0.133 &	28.401  $\pm$  0.168	& 72.78/20  \\\hline

172 	& 4.537  $\pm$  0.307	& 24.380  $\pm$  0.691 &	5.363  $\pm$  0.963 &	34.070  $\pm$  0.928	& 7.83/23  \\\hline

183 	& 5.025  $\pm$  0.139	& 25.170  $\pm$  0.310 &	5.157  $\pm$  0.358 &	37.510  $\pm$  0.516	& 28.80/23	 \\\hline

189 	& 4.629  $\pm$  0.094 &	25.570  $\pm$  0.260 &	5.121  $\pm$  0.377 &	36.690  $\pm$  0.405	& 52.11/23	 \\\hline

& 	  			 	 &	&	 &	  &   \\ 
						
130.1 	& 4.811  $\pm$  0.248	 & 22.267  $\pm$  0.469 &	4.696  $\pm$  0.352 &	31.041  $\pm$  0.438	& 9.751/17	 \\\hline

136.1	 & 4.286  $\pm$  0.318	& 22.681  $\pm$  0.858 &	4.808  $\pm$  0.698 &	32.110  $\pm$  0.575 &	27.63/17  \\\hline

172.3	 & 5.054  $\pm$  0.218 &	25.191  $\pm$  0.624 &	4.631  $\pm$  0.318 &	37.133  $\pm$  0.602 &	11.26/17	 \\\hline

182.8 	& 4.757  $\pm$  0.140	& 24.91  $\pm$  0.322	& 5.315  $\pm$  0.497 &	36.765  $\pm$  0.570	& 29.92/17	 \\\hline

188.6 	& 4.721  $\pm$  0.096	& 25.051  $\pm$  0.184 & 5.445  $\pm$  0.326 & 36.492  $\pm$  0.422	& 40.93/17  \\\hline

194.4	 & 4.695  $\pm$  0.131 &	25.454  $\pm$  0.299 &	4.659  $\pm$  0.300 &	37.322  $\pm$  0.601	& 30.86/17	 \\\hline

200.2 	& 4.915  $\pm$  0.120 &	25.473  $\pm$  0.248	& 5.197  $\pm$  0.398	& 38.391  $\pm$  0.670	& 29.01/17	 \\\hline

206.2 	 & 4.846  $\pm$  0.106 &	25.582  $\pm$  0.226 &	5.602  $\pm$  0.437 &	38.210  $\pm$  0.564	& 41.41/17	 \\\hline

\end{tabular}
\caption{Parameters of Modified Weibull function for OPAL $\&$ L3 data.}
\end{table*}

\begin{table*}[t]
\small
\begin{tabular}{|c|c|c|c|c|c|c|c|c|}
\hline
Energy & $nV_{1}$ & $nv_{0_1}$ & $nV_{2}$ &$nv_{0_2}$  & $K_{1}$ & $K_{2}$  \\
 	& &	  		 &		&		&	 &	     \\\hline    
91  &	10.080  $\pm$  5.942	& 0.444  $\pm$  0.227	& 10.390  $\pm$  8.953	& 0.415  $\pm$  0.192	& 22.140  $\pm$  2.022	 & 50.001  $\pm$  3.202\\\hline 
131  &	14.180  $\pm$  2.312	& 0.121  $\pm$  0.322	& 16.760  $\pm$  4.280	& 0.254  $\pm$  0.260	& 68.540  $\pm$  31.590	& 46.340  $\pm$  37.620\\\hline 
161  &	14.450  $\pm$  1.746	& 0.176  $\pm$  0.258	& 13.190  $\pm$  3.718	& 0.124  $\pm$  0.280	& 24.890  $\pm$  5.287	& 50.001  $\pm$  35.600\\\hline 
172  &	11.350  $\pm$  0.952	& 0.409  $\pm$  0.168	& 19.920  $\pm$  5.090	& 0.153  $\pm$  0.222	& 33.920  $\pm$  11.590	& 50.002  $\pm$  36.680\\\hline 
183  &	12.201  $\pm$  0.489	& 0.401  $\pm$  0.028	& 11.130  $\pm$  0.219	& 0.445  $\pm$  0.003	& 17.970  $\pm$  2.197	& 50.001  $\pm$  3.275\\\hline 
189  &	11.080  $\pm$  0.265	& 0.404  $\pm$  0.042	& 15.880  $\pm$  1.716	& 0.387  $\pm$  0.033	& 32.101  $\pm$  5.396	&37.410  $\pm$  12.660\\\hline 

& &	  &				&		&	 &	     \\						
130.1  &	14.011  $\pm$  4.502	& 0.134  $\pm$  0.174	& 13.741  $\pm$  1.041	& 0.181  $\pm$  0.224	& 25.032  $\pm$  4.022	& 50.027  $\pm$  28.561\\\hline 
136.1  &	14.158  $\pm$  1.183	& 0.119  $\pm$  0.325	& 11.433  $\pm$  3.861	& 0.303  $\pm$  0.245	& 24.591  $\pm$  3.979	& 50.021  $\pm$  27.911\\\hline 
172.3  &	22.991  $\pm$  0.336	& 0.122  $\pm$  0.016	& 12.511  $\pm$  0.095	& 0.352  $\pm$  0.009	& 33.573  $\pm$  5.151	& 50.007  $\pm$  38.722\\\hline 
182.8  &	22.363  $\pm$  4.952	& 0.167  $\pm$  0.020	& 12.343  $\pm$  0.989	& 0.332  $\pm$  0.093	& 31.388  $\pm$  3.787	& 50.001  $\pm$  27.303\\\hline 
188.6   &	18.811  $\pm$  3.114	& 0.305  $\pm$  0.072	& 12.733  $\pm$  0.761	& 0.337  $\pm$  0.263	& 30.031  $\pm$  2.654	& 50.004  $\pm$  8.850\\\hline 
194.4   &	19.435  $\pm$  0.352	& 0.303  $\pm$  0.010	& 12.770  $\pm$  0.111	& 0.243  $\pm$  0.018	& 31.455  $\pm$  3.500	& 50.006  $\pm$  37.741\\\hline 
200.2   &	24.487  $\pm$  5.641	& 0.312  $\pm$  0.075	& 12.881  $\pm$  0.843	& 0.115  $\pm$  0.228	& 34.391  $\pm$  3.687	& 50.005  $\pm$  30.089\\\hline 
206.2  &	20.030  $\pm$  3.222	& 0.333  $\pm$  0.245	& 15.191  $\pm$  0.464	& 0.115  $\pm$  0.005	& 33.101  $\pm$  2.983	& 50.012  $\pm$  28.161\\\hline 

\end{tabular}
\caption{Parameters of Modified Tsallis function for  OPAL $\&$ L3  data.}
\end{table*}

\begin{table*}[t]
\small
\begin{tabular}{|c|c|c|c|}
\hline
Energy   & $q_{1}$ & $q_{2}$ &  $\chi^{2}/ndf$ \\
 	& &	  		&      \\\hline  

91 	& 1.058  $\pm$  0.153	& 1.577  $\pm$  0.776	& 5.32/16	\\\hline
133	& 1.023  $\pm$  0.129	& 1.120  $\pm$  0.541	& 2.40/16	\\\hline
161 	& 1.065  $\pm$  0.160	& 1.002  $\pm$  0.551	& 2.89/16	\\\hline
172 	& 1.210  $\pm$  0.229	& 1.799  $\pm$  0.838	& 3.31/19	\\\hline
183 	& 1.086  $\pm$  0.065	& 1.061  $\pm$  0.059	& 12.91/19	\\\hline
189 	& 1.005  $\pm$  0.013	& 1.112  $\pm$  0.097	& 2.73/19	\\\hline
& &	  		&     \\				
130.1 	& 1.026  $\pm$  0.019	& 1.988  $\pm$  0.887	& 4.20/13	\\\hline
136.1 	& 1.001  $\pm$  0.025	& 1.772  $\pm$  0.539	& 16.61/13   \\\hline
172.3 	& 1.187  $\pm$  0.052	& 1.498  $\pm$  0.232	& 1.51/13	\\\hline
182.8 	& 1.089  $\pm$  0.171	& 1.958  $\pm$  0.921	& 5.72/13	\\\hline
188.6 	& 1.147  $\pm$  0.154	& 1.789  $\pm$  0.680	& 5.83/13	\\\hline
194.4 	& 1.086  $\pm$  0.037	& 1.001  $\pm$  0.100	& 6.94/13	\\\hline
200.2 	& 1.065  $\pm$  0.151	& 1.825  $\pm$  0.767	& 4.61/13	\\\hline
206.2 	& 1.024  $\pm$  0.060	& 1.331  $\pm$  0.364	& 4.12/13	\\\hline

\end{tabular}
\caption{q values of Modified Tsallis function for OPAL $\&$ L3 data.}
\end{table*}

\begin{table*}[t]
\begin{tabular}{|c| c|c |c|c| c|c| c|c| c|c|}
\hline
Energy  & Weibull &  & Tsallis &  & Modified  &  & Modified    &    \\
(GeV)&       & 		 & 		& 		& 	Weibull	& 		& 		Tsallis 	   &   	 \\\hline  

& $\chi^2/ndf$ & p value & $\chi^2/ndf$  & p value & $\chi^2/ndf$  & p value & $\chi^2/ndf$  & p value  \\
&       & 		 & 		& 		& 		& 		& 		& 		  	 \\\hline

91 &	434.91/22 &	0.0001	 & 33.13/20	 & 0.0329	 & 12.20/20	 & 0.9090	 & 5.32/16	 & 0.9940\\\hline
133 &	66.88/22  &	0.0001	 & 12.67/20	 & 0.8899	 & 14.71/20	 & 0.7933	& 2.40/16	 & 1.0000\\\hline
161 &	 48.11/22  &	0.0011	 & 5.81/20	 & 0.9991	 & 72.78/20	 & 0.0001 &	2.89/16	 & 0.9999\\\hline 
172 &	17.87/22  &	0.8466	 & 3.94/20	 & 1.000	 	& 7.83/23	 & 0.9987	& 3.31/19	 & 1.0000\\\hline
183 &	159.61/25  &	0.0001	 & 35.13/23	 & 0.0508	 & 28.80/23	 & 0.187	& 12.91/19	 & 0.8437\\\hline
189 &	183.63/25 &	0.0001	 & 15.88/23	 & 0.8595	 & 52.11/23	 & 0.0005 &	2.73/19	 & 1.0000\\\hline  
&       & 		 & 		& 		& 		& 		& 		& 		  	 \\
130.1	& 52.53/19	& 0.0001 	& 6.42/17	& 0.9899	& 9.751/17 	& 0.9137	& 4.20/13	& 0.9889\\\hline
136.1	& 41.64/19	& 0.002		& 19.26/17	& 0.3138	& 27.63/17	& 0.0495	& 16.61/13	& 0.2178\\\hline
172.3	& 64.28/19	& 0.0001	& 8.78/17	& 0.947		& 11.26/17	& 0.8427	& 1.51/13	& 1.0000\\\hline
182.8	& 98.71/19	& 0.0001	& 16.98/17	& 0.4557	& 29.92/17	& 0.0269	& 5.72/13	& 0.9558\\\hline
188.6	& 157.54/19	& 0.0001	& 17.28/17	& 0.4356	& 40.93/17	& 0.001		& 5.83/13	& 0.9521\\\hline
194.4	& 108.70/19	& 0.0001	& 19.19/17	& 0.3177	& 30.86/17	& 0.0208	& 6.94/13	& 0.9052\\\hline
200.2	& 127.53/19	& 0.0001	& 27.14/17	& 0.056		& 29.01/17	& 0.0344	& 4.61/13	& 0.9828\\\hline
206.2	& 168.91/19	& 0.0001	& 32.41/17	& 0.0134	& 41.41/17	& 0.0008	& 4.12/13	& 0.9898\\\hline

\end{tabular}
\caption{$\chi^{2}/ndf$ comparison $\&$ p values for different energies of OPAL $\&$ L3 experiment for Weibull, Tsallis,Modified Tsallis  $\&$ modified Weibull distibutions} 
\end{table*}

\end{document}